\documentclass[12pt]{iopart}

\expandafter\let\csname equation*\endcsname\relax
\expandafter\let\csname endequation*\endcsname\relax

\usepackage{graphicx,color}
\usepackage{amsmath,amssymb,latexsym}
\usepackage{mathtools}
\usepackage[english]{babel}
\usepackage{dcolumn}
\usepackage{bm}
\usepackage{float}
\usepackage[normalem]{ulem}

\usepackage{color}

\begin{document}

\title[Brownian non-Gaussian diffusion of self-avoiding walks]{Brownian non-Gaussian diffusion of self-avoiding walks}

\author{Boris Marcone, Sankaran Nampoothiri, Enzo Orlandini, Flavio Seno, Fulvio Baldovin}
\address{Dipartimento di Fisica e Astronomia `G. Galilei' - DFA, Sezione INFN,
Universit\`a di Padova,
Via Marzolo 8, 35131 Padova (PD), Italy} 
\ead{marcone@pd.infn.it}
\ead{sankaran.nampoothiri@unipd.it}
\ead{enzo.orlandini@unipd.it}
\ead{flavio.seno@unipd.it}
\ead{fulvio.baldovin@unipd.it}





\date{\today}

\begin{abstract}
  Three-dimensional Monte Carlo simulations provide a striking
  confirmation to a recent theoretical prediction: the Brownian
  non-Gaussian diffusion of critical self-avoiding walks.  Although
  the mean square displacement of the polymer center of mass grows
  linearly with time (Brownian behavior), the initial probability
  density function is strongly non-Gaussian and crosses over to
  Gaussianity only at large time.  Full agreement between theory and
  simulations is achieved without the employment of fitting
  parameters. We discuss simulation techniques potentially capable of
  addressing the study of anomalous diffusion under complex conditions
  like adsorption- or Theta-transition.
\end{abstract}

\submitto{\JPA}
\maketitle

\section{Introduction}
A wealth of recent
experiments~\cite{wang2009,wang2012,toyota2011,yu2013,yu2014,chakraborty2020,weeks2000,wagner2017,jeon2016,yamamoto2017,stylianidou2014,parry2014,munder2016,cherstvy2018,li2019,cuetos2018,hapca2008,pastore2021rapid}
  and molecular dynamics simulations~\cite{pastore2015,miotto2021length,pastore2022}
highlighted the existence of diffusion processes in which the mean
square displacement linearly grows in time (in agreement with Brownian
behavior), but the probability density function (PDF) of the tracer
displacement is non-Gaussian during long stages.  This has triggered
interest for the introduction of novel mesoscopic models relying on
superposition of
statistics~\cite{beck2003,beck2006,hapca2008,wang2012}, diffusing
diffusivities~\cite{chubynsky2014,chechkin2017,jain2017,
  tyagi2017,miyaguchi2017, sposini2018,
  sposini2018first}, continuous time random
walk~\cite{barkai2020,wang2020,sokolov2021ld}, and diffusion in
disordered environments~\cite{sokolov2021}, as well as theoretical
approaches based on a clear microscopic
foundation~\cite{baldovin2019,nampoothiri2021,nampoothiri2022,hidalgo2020}.
A simple microscopic prototype displaying Brownian non-Gaussian 
diffusion is represented by a tracer undergoing conformational
modifications which affect its diffusion coefficient. Consider for
instance the centre of mass (CM) of a polymer in contact with a
chemostat, so that its size $N$ may fluctuate as it does in the grand
canonical ensemble.  Since the CM diffusion coefficient is a function
of the size, size fluctuations naturally generate diffusivity
fluctuations and this renders the initial PDF of the CM displacement
non-Gaussian (see below). Importantly, the phenomenon becomes evident
close to the
polymer critical point separating the dilute from the dense
phase~\cite{deGennes1972,deGennes1979,vanderzande1998,madras2013},
where size 
fluctuations diverge and the behaviour
becomes universal, only depending on the spatial dimension $d$, on the
polymer topology, and the solvent
conditions.
This state of affairs has been recently theorised
in Refs.~\cite{nampoothiri2021,nampoothiri2022}, where explicit predictions
about the initial  PDF shape have been made based on the critical exponents
characterising the polymer universality class.

By exploiting the fact that a microscopic model is endowed with its
own independent dynamics, here we consider grand canonical (i.e. $N$
varying) simulations of the three-dimensional ($d=3$) self-avoiding
walks (SAW) on the cubic lattice as a model for a linear polymer
fluctuating in size under good solvent
conditions~\cite{vanderzande1998}.
The use of Monte Carlo SAW schemes to simulate polymer dynamics dates back
to the work of Baumgartner and Binder~\cite{baumgartner1979}, where it
was implemented the so-called ``kink-jump dynamics''.
Since then, many processes involving polymer
dynamics have been simulated via the use of Monte Carlo
methods (see for instance Refs.~\cite{chern2001,burroughs2007}).
Although these schemes do not fully mimic Netwonian dynamics
-- e.g., they disregards inertial effects -- in many cases they
faithfully describe the diffusion properties of the
system and allow for massive statistical sampling with respect to the
molecular dynamics counterparts.
The algorithm outlined in Section 3, based on a variant of the
``kink-jump'' suitable for SAWs on a cubic lattice,
will be shown to correctly reproduce
the polymer diffusion dynamics.
After reviewing the
universality class of three-dimensional SAWs, we address the CM
displacements during the simulation dynamics and show a striking
agreement between the corresponding PDF and the theoretical
predictions reported in
Refs.~\cite{nampoothiri2021,nampoothiri2022}. Importantly, this
conformity is achieved without employing any fitting parameter and
provides a standalone microscopic validation for the Brownian
non-Gaussian character of the CM motion of polymers in proximity of
their critical point.

The paper is organised as follows. In the next Section we provide the
theoretical background for three-dimensional SAWs and 
for the diffusion properties of their CM. We then supply details of
the simulation methods and end the paper by a discussion of our results and
future developments.

\section{Theoretical background}
SAWs are a fundamental mathematical model employed to reproduce the
properties of linear polymers in good solvent~\cite{vanderzande1998}.
Assuming for simplicity the same energy for all SAWs, their
equilibrium statistical behaviour in the grand canonical ensemble is described by the 
partition (or generating) function
\begin{equation}
  Z_{\mathrm{gc}}(z)
  =\sum_n  c_n\,z^n\,,
  =\sum_\omega  z^{|\omega|}\,,
\end{equation}
where $c_n$ is the number of distinct $n$-step SAWs, and $z$ the step
fugacity; the second sum is over all SAWs $\omega$, of arbitrary
size $|\omega|$.
In the grand canonical ensemble, the fugacity
$z=\mathrm{e}^{\mu/k_{\mathrm{B}}T}$ (where $\mu$ is the chemical
potential, $T$ temperature, and $k_{\mathrm{B}}$ the Boltzmann
constant) corresponds to an intensive
thermodynamic variable controlling the interchanges of particles between the
system and the chemostat; for SAWs, added and removed particles
amount to added and removed steps.
It is well known that a model-dependent critical fugacity
$z_{\mathrm{c}}$ separates the dilute polymer phase in which the
average size $\mathbb{E}[N]$ is finite from a dense one characterized
by a divergent $\mathbb{E}[N]$; with a three-dimensional cubic lattice
critical fugacity, $z_{\mathrm{c}}
\simeq0.213490$~\cite{clisby2013calculation}.
As $z\to
z_{\mathrm{c}}^-$ the partition function takes the universal form
\begin{equation}
  Z_{\mathrm{gc}}(z)=\sum_{n=n_{\mathrm{min}}}^\infty
  (z/z_{\mathrm{c}})^n\,n^{\gamma-1}
  =\mathrm{Li}_{1-\gamma}(z/z_{\mathrm{c}})
  -\sum_{n=1}^{n_{\mathrm{min}}}(z/z_{\mathrm{c}})^n\,n^{\gamma-1}
  \sim(1-z/z_{\mathrm{c}})^{-\gamma}\,,
  \label{grancan_scal}
\end{equation}
where $\gamma$ is the critical entropic exponent whose estimate in
$d=3$ is $\gamma=1.156953 00 (95)$ \cite{clisby2017scale},
and $\mathrm{Li}_s(z)\equiv\sum_{n=1}^\infty z^n/n^s$ the
polylogarithm function. In Eq.~\eqref{grancan_scal} we have assumed a
minimal number of steps $n_{\mathrm{min}}$, which will be discussed
below. 

\begin{figure}[t]
  \begin{center}
    \includegraphics[width=0.48\columnwidth]{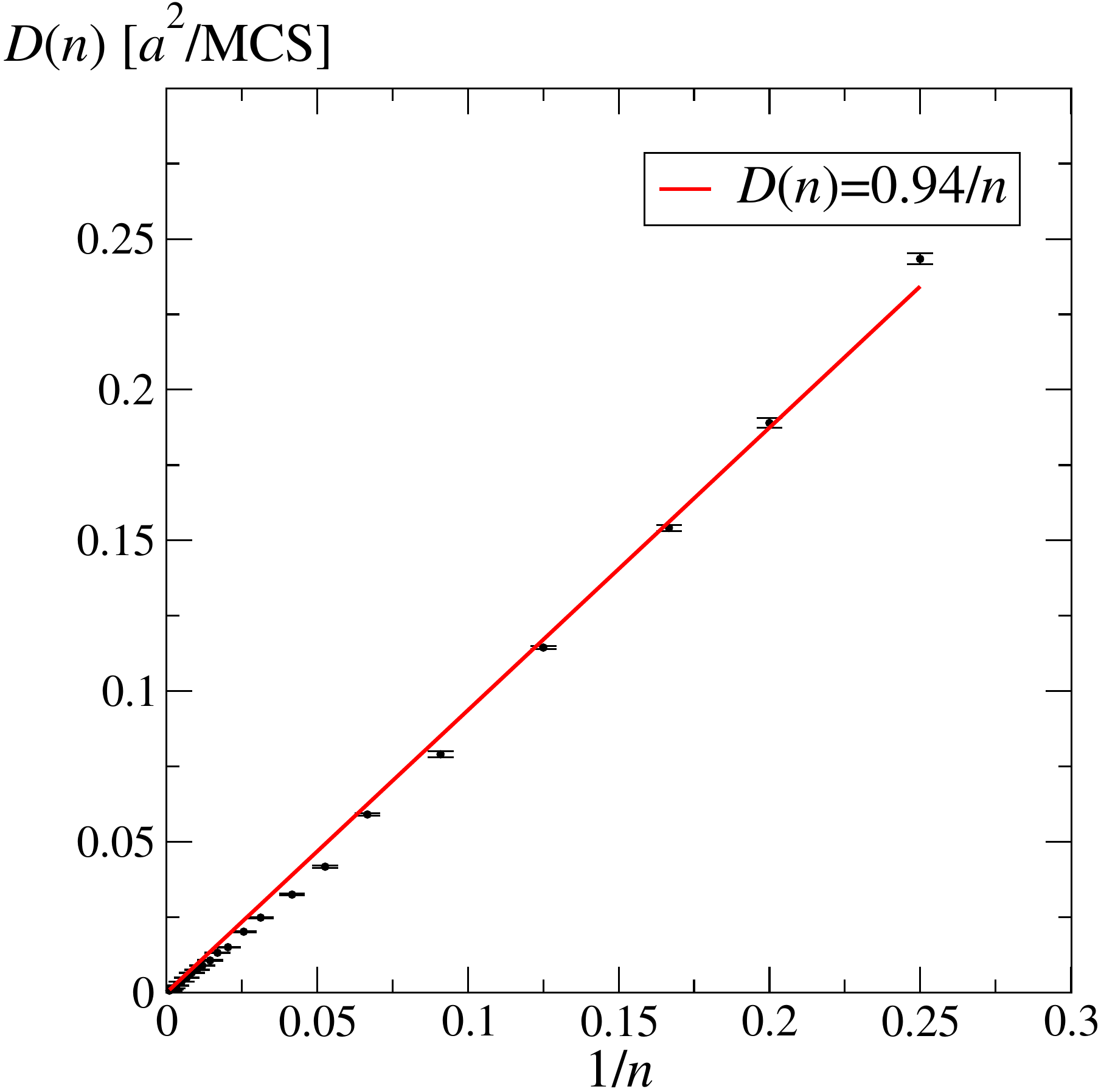}
    \includegraphics[width=0.48\columnwidth]{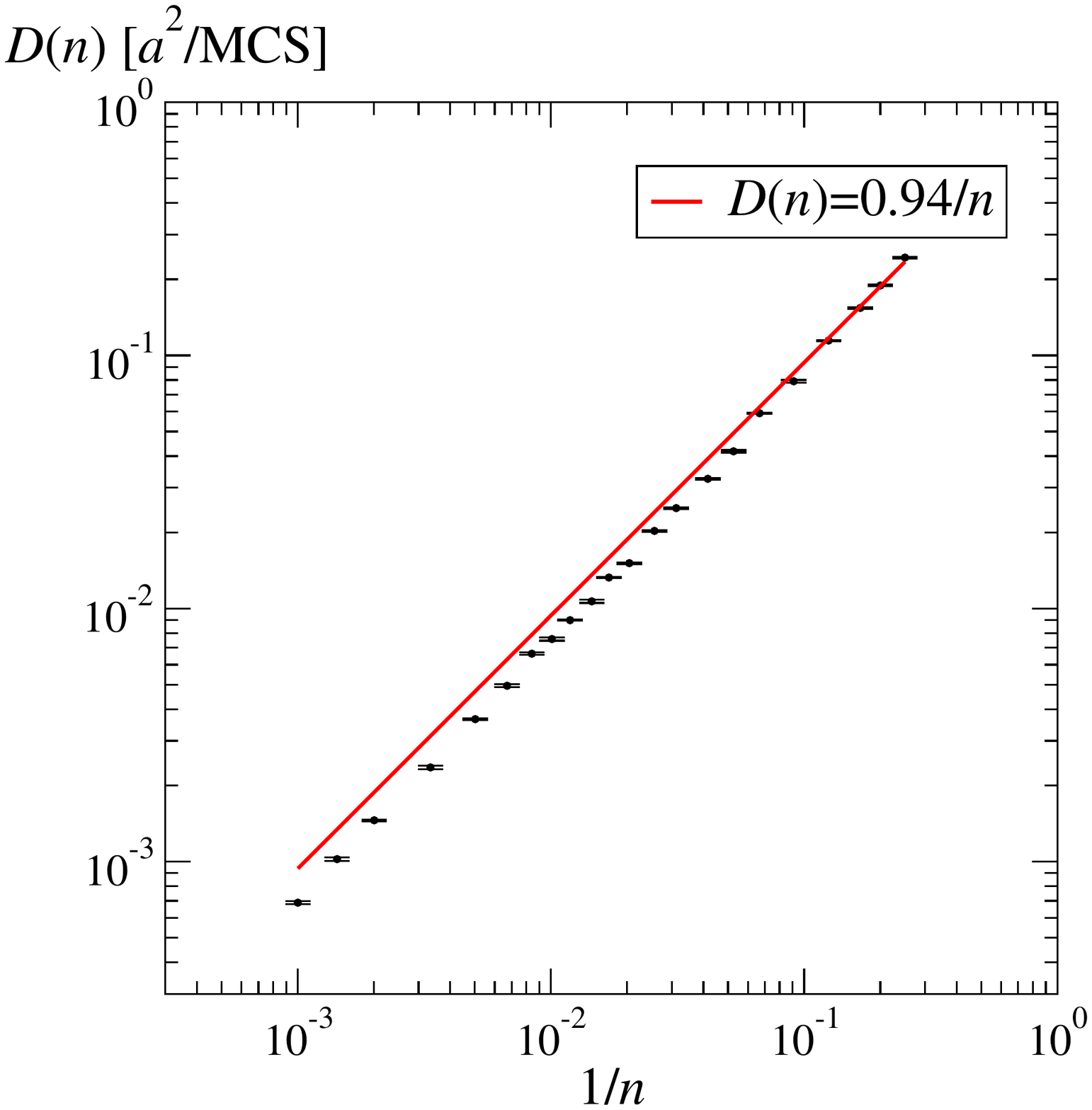}\\
    \caption{CM diffusion coefficient for fixed size $n\geq4$ SAWs in a
      cubic lattice of constant $a$,
      both in linear and log-log scale.
      Symbols:
      $D(n)$ is measured in units of $a^2$ over Monte Carlo Steps
      (MCS) -- see Section~\ref{sec_motion_cm} for the description of 
      $n$-preserving Monte Carlo moves. Line: null-intercept linear
      regression yielding $D_{\mathrm{m}}=0.94\;a^2/\mathrm{MCS}$ in
      Eq.~\eqref{eq_rouse}, with
      regression coefficient $r=0.9991$.}
    \label{fig_d_vs_1on}
  \end{center}
\end{figure}

If we now consider SAWs  of fixed size $N=n\geq n_{\mathrm{min}}$,
from one hand Eq.~\eqref{grancan_scal} tells that their equilibrium
occurrence is given by the probability distribution
\begin{equation}
  P_N^*(n)
  =\dfrac{(z/z_{\mathrm{c}})^n\;n^{\gamma-1}
  }{Z_{\mathrm{gc}}(z)}
  \sim(1-z/z_{\mathrm{c}})^{\gamma}\;(z/z_{\mathrm{c}})^n\;n^{\gamma-1}\,.
\label{scaling_Pn}
\end{equation}
On the other hand, random conformational changes of these SAWs  induce
a diffusive motion for the position
$\boldsymbol{R}_{\mathrm{CM}}(t)=(X_{\mathrm{CM}}(t),Y_{\mathrm{CM}}(t),Z_{\mathrm{CM}}(t))$
of their CM according to 
\begin{equation}
\mathrm{d}\boldsymbol{R}_{\mathrm{CM}}(t)=\sqrt{2D(n)}
\,\mathrm{d}\boldsymbol{B}(\mathrm{d}t)\,,
\label{eq_langevin}
\end{equation}
with $\boldsymbol{B}(\mathrm{d}t)$ being a Wiener process (Brownian
motion) and $D(n)$ the corresponding diffusion coefficient.  As
explained in the next Section, the stochastic motion of the SAWs is
implemented via a kinetic Monte Carlo with local deformations on the
lattice.  As a consequence, hydrodynamic effects are not taken into
account by our dynamics and the size-dependency of the diffusion
coefficient is expected to be Rouse-like~\cite{Doi1992}, namely
\begin{equation}
  D(n)\simeq\dfrac{D_{\mathrm{m}}}{n}\,,
  \label{eq_rouse}
\end{equation}
for sufficiently large $n$'s and $D_{\mathrm{m}}$ being an effective
single-monomer diffusion coefficient.
Figure~\ref{fig_d_vs_1on} confirms this expectation. 

Coming back to the grand canonical ensemble, size fluctuations may be
described in terms of a birth-death process
$N(t)$~\cite{nampoothiri2021,nampoothiri2022}:
\begin{equation}
  \begin{array}{ll}
    \partial_t P_N(n,t|n_0)
    =&
    \mu\,P_N(n+1,t|n_0)
  +\lambda(n-1)\,P_N(n-1,t|n_0)
    \vspace{0.1cm}
    \\
    &-(\mu+\lambda(n))P_N(n,t|n_0)
    \qquad\qquad\;\;\;\;(n>n_{\mathrm{min}})
    \vspace{0.1cm}
    \\
    \partial_t P_N(n_{\mathrm{min}},t|n_0)
    =&
    \mu\,P_N(n_{\mathrm{min}}+1,t|n_0)
    -\lambda(n_{\mathrm{min}})\,P_N(n_{\mathrm{min}},t|n_0)\,
  \end{array}
  \label{eq_master}
\end{equation}
where $\lambda(n)$ and $\mu$ are respectively the  association and
dissociation rates.
Here, $P_N(n,t|n_0)$ is the probability for $N(t)=n$, given
$N(0)=n_0$.
Calling $\tau$ the autocorrelation time of $N(t)$, the expected asymptotic
behaviours for $P_N(n,t|n_0)$ are
\mbox{$P_N(n,t|n_0)\substack{\sim\\t\ll\tau}\delta_{n,n_0}$} and
\mbox{$P_N(n,t|n_0)\substack{\sim\\t\gg\tau}P^\star_N(n)$}.
Note that we have assumed the dissociation rate to be
independent of the size $n$, since it typically represents a decay
process occurring at the extremities of the  chain. Conversely, the detail
balance condition applied to the equilibrium distribution
straightforwardly yields~\cite{nampoothiri2022}
\begin{equation}
  \lambda(n)
  =\mu\,\dfrac{P^\star_N(n+1)}{P^\star_N(n)}
  =\mu\,\dfrac{z}{z_{\mathrm{c}}}\,\left(\dfrac{n+1}{n}\right)^{\gamma-1}\,.
\end{equation}
In what follows $\mu$, with the dimension of an inverse time, can be
regarded as a free parameter which may be tuned to match the model
with the autocorrelation $\tau$ of real physical
conditions~\cite{nampoothiri2022}. 

A fluctuating $N(t)$ adds a second source of randomness 
to Eq.~\eqref{eq_langevin}; in the literature, $N(t)$ is called the
\textit{subordinator} and $\boldsymbol{B}(t)$ the
\textit{subordinated} process~\cite{Feller1968,bochner2020harmonic}. 
It is convenient to reparametrize the diffusion path in
terms of the coordinate $s\geq0$,
$\mathrm{d}s=2\,D(n(t))\,\mathrm{d}t$, corresponding to the
realization of the stochastic process
\begin{equation}
  S(t)\equiv2\int_0^t\mathrm{d}t'\,D(N(t'))
  =2D_{\mathrm{m}}\int_0^t\mathrm{d}t'\,N^{-1}(t')\,,
\end{equation}
with PDF $p_S(s,t|n_0)$.
The conditional PDF for the CM displacement
$\boldsymbol{R}_{\mathrm{CM}}=\boldsymbol{r}$ at time 
$t$ given the initial conditions $N=n_0$,
$\boldsymbol{R}_{\mathrm{CM}}=\boldsymbol{0}$ at time $0$,  
$p_{\boldsymbol{R}_{\mathrm{CM}}}(\boldsymbol{r},t|n_0;\boldsymbol{0})$,
is then obtained through the subordination
formula~\cite{chechkin2017,nampoothiri2021,nampoothiri2022}
\begin{equation}
  p_{\boldsymbol{R}_{\mathrm{CM}}}(\boldsymbol{r},t|n_0;\boldsymbol{0})
  =\int_0^\infty\mathrm{d}s
  \,\dfrac{\mathrm{e}^{-\frac{\boldsymbol{r}^2}{2s}}}{(2\pi\,s)^{3/2}}
  \;p_S(s,t|n_0)\,.
  \label{eq_subordination}
\end{equation}
Taking an equilibrium distribution $P^\star_N(n_0)$ for the initial
sizes of the SAW, from Eq.~\eqref{eq_subordination} one gets the
Brownian character of the CM diffusion:
\begin{equation}
  \mathbb{E}\left[\boldsymbol{R}_{\mathrm{CM}}^2(t)\right]
  =3\,\mathbb{E}\left[S(t)\right]=6\,D_{\mathrm{av}}\,t\,,
\end{equation}
with
$D_{\mathrm{av}}\equiv\displaystyle\sum_{n=n_{\mathrm{min}}}^\infty\dfrac{D_{\mathrm{m}}}{n}\,P^\star_N(n)$.  
The shape of the initial non-Gaussian PDF for the polymer CM is
instead conveniently studied by switching to the unit-variance
dimensionless variable
$\overline{X}_{\mathrm{CM}}(t)\equiv X_{\mathrm{CM}}(t)/\sqrt{\mathbb{E}[X_{\mathrm{CM}}^2(t)]}$.
From Eq.~\eqref{eq_subordination} and the asymptotic
limits of the birth-death process, as $t\to0^+$ we have 
\begin{equation}
  p_{\overline{X}_{\mathrm{CM}}}(x,0^+)
  \simeq\sum_{n=n_{\mathrm{min}}}^\infty P^\star_N(n)
  \,\dfrac{\mathrm{e}^{-\frac{\mathbb{E}[N^{-1}]\,x^2}{2\,n^{-1}}}
  }{
    \sqrt{2\pi\,\frac{n^{-1}}{\mathbb{E}[N^{-1}]}}
  }\,.
  \label{eq_initial_pdf}
\end{equation}
At large $|x|$ the PDF is
asymptotic to the Gaussian cutoff
$\sim\mathrm{e}^{-\mathbb{E}[N^{-1}]\,x^2/(2\,n_{\mathrm{min}}^{-1})}$, and as
$z/z_{\mathrm{c}}\to1$ 
this cutoff is pushed towards $|x|\to\infty$, since
$\mathbb{E}[N^{-1}]\to0$.
Conversely, in the long time limit the ordinary Gaussian behavior
is restored, $p_{\overline{X}_{\mathrm{CM}}}(x,\infty)\sim\mathrm{e}^{-x^2/2}$. The
autocorrelation $\tau$ is the time scale separating  the two behaviours;
as $z\to z_{\mathrm{c}}^-$ this
autocorrelation time diverges~\cite{nampoothiri2022} (critical slowing
down).  

\begin{figure}[t]
  \begin{center}
    \includegraphics[width=0.48\columnwidth]{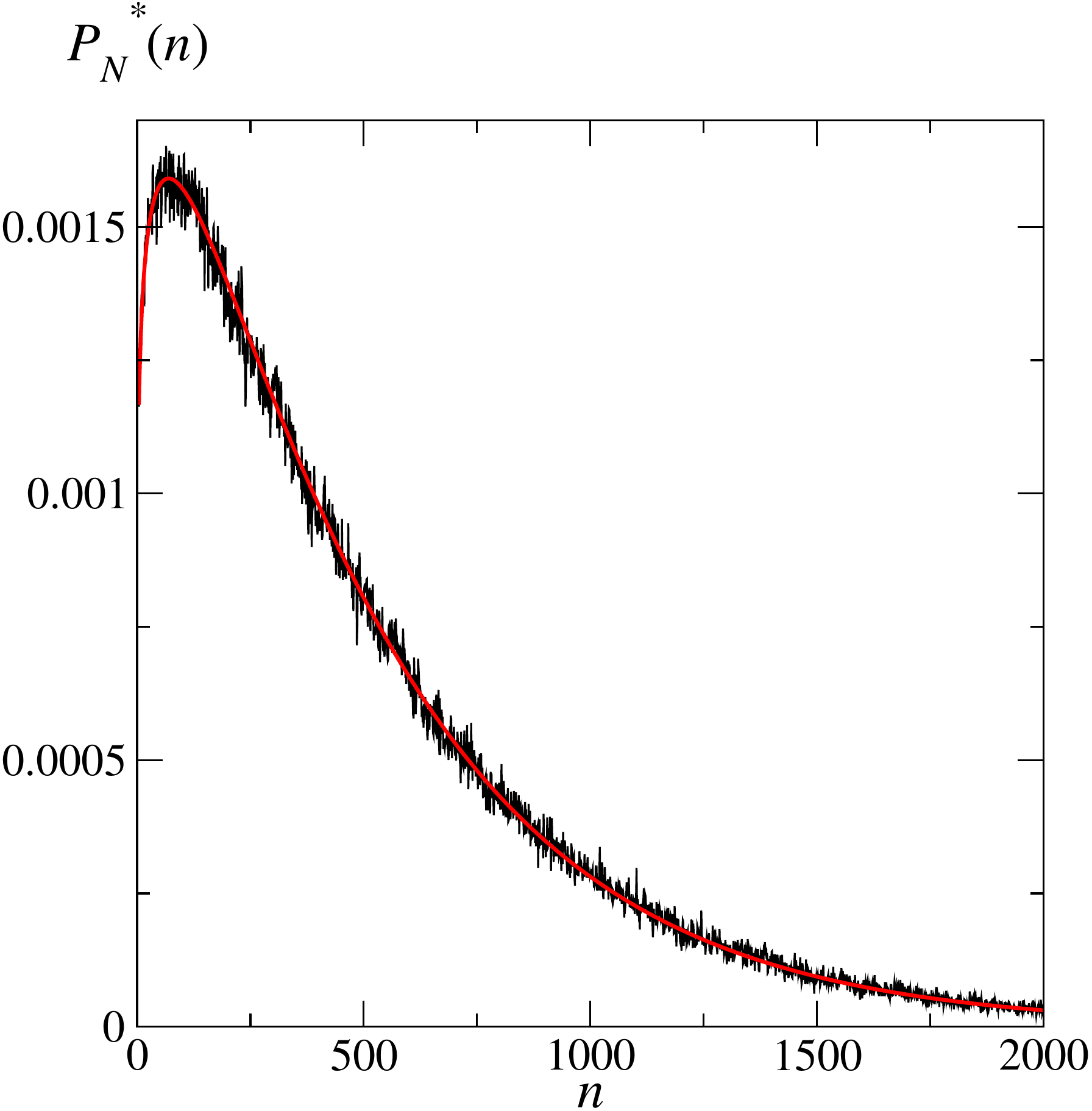}
    \includegraphics[width=0.48\columnwidth]{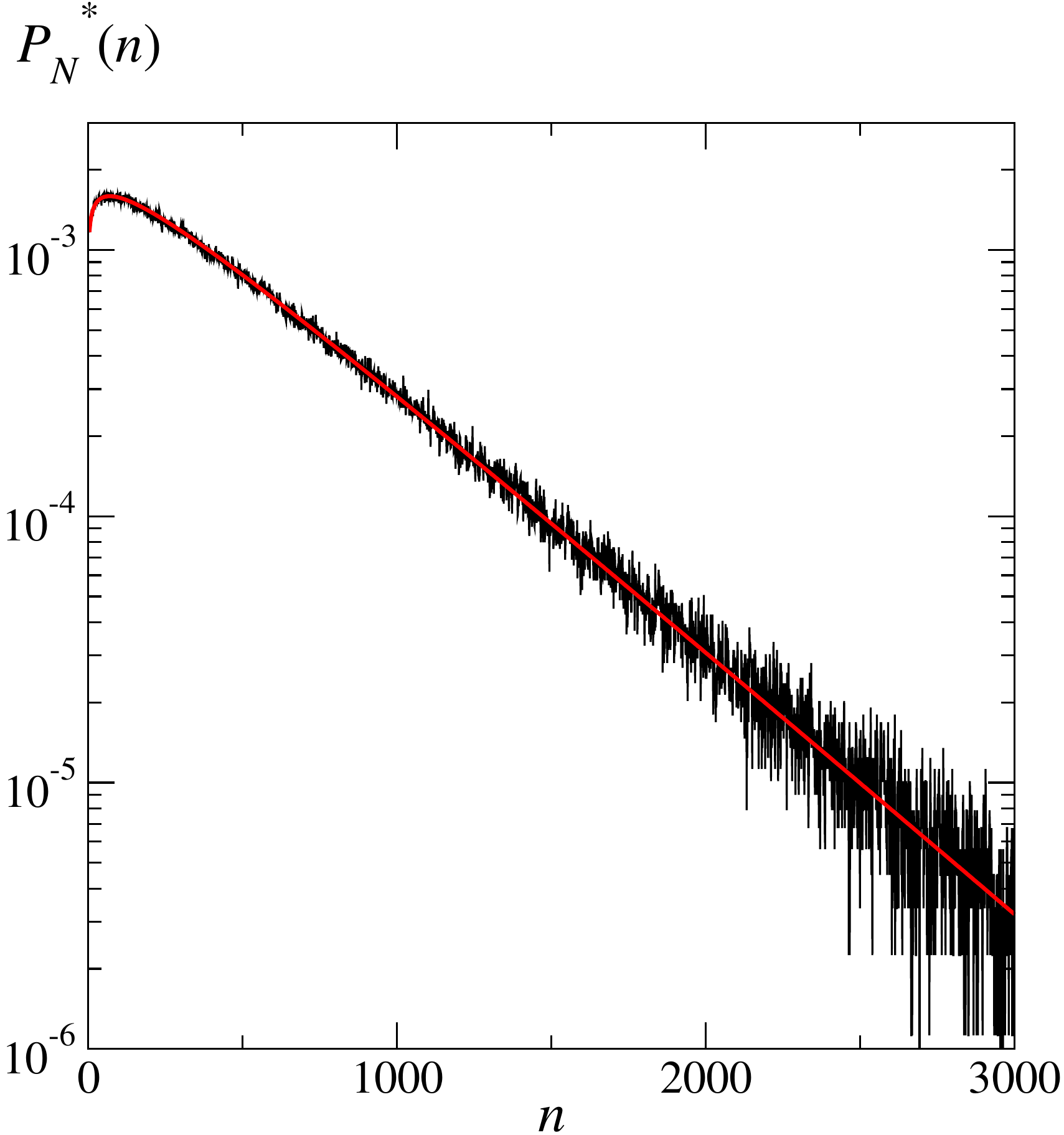}\\
    \caption{Comparison between the Monte Carlo estimation for
      $P_N^\star(n)$ (black line) and Eq.~\eqref{scaling_Pn} (red
      line), both in linear and log-linear plot.} 
    \label{fig_p_n}
  \end{center}
\end{figure}

\section{Simulation methods}
\label{sec_methods}
On-lattice models are a schematic representation of actual polymers,
allowing for massive data sampling. In the present simplification in
which all configurations share the same energy, they are particularly
effective in reproducing the equilibrium statistics under detail
balance conditions.
Since we want to sample SAWs in the grand canonical ensemble, we
apply a Monte Carlo method based on a variation of the
Berretti-Sokal algorithm~\cite{berretti1985new}, perhaps the simplest
variable-length dynamic Monte Carlo method whose state space is the
set of all SAWs (i.e. SAWs on any lengths) in a cubic lattice. The basic idea of the
original implementation is that, at each iteration, one of the two ends of
the chain is randomly chosen; then, with probability
$P_N(n-1|n)$ the last step of the walk is  deleted (deletion or $-1$
move), while  with probability $P_N(n+1|n)=1-P_N(n-1|n)$ an attempt is
made to increase the SAWs length by appending a new edge with equal 
probability in each of the $2d=6$ possible directions ($+1$ move).
In the latter case, if the new configuration is not self-avoiding the proposed move is
rejected and the old configuration is counted again in the sampling
(null transition).  A null transition is also made if a deletion move
is attempted on a walk with $n=n_{\mathrm{min}}$ steps.
To assure that the invariant probability distribution of a walk is
correctly given by  
$P_\Omega^\star(\omega)=z^{|\omega|}/Z_{\mathrm{gc}}(z)$,
the transition probabilities for the $\pm1$ moves are taken according
to~\cite{berretti1985new}  
\begin{equation}
  P_N(n+1|n) = \frac{2dz}{1+2dz},\qquad
  P_N(n-1|n) = \frac{1}{1+2dz}.
\end{equation}
The implementation of this algorithm recovers the estimates of
$z_{\mathrm{c}}$ and $\gamma$ 
given above; in particular, Figure~\ref{fig_p_n} reports a very good
agreement between the equilibrium size distribution sampled with the
Monte Carlo strategy and Eq.~\eqref{scaling_Pn}.
Grand canonical Monte Carlo simulations reported in Figure~\ref{fig_p_n}
and in the following ones are performed at
$z/z_{\mathrm{c}}=0.997677$; evaluation of the autocorrelation
time for this fugacity returns $\tau\simeq\times10^6\;\mathrm{MCS}$.

\begin{figure}[t]
  \begin{center}
    \includegraphics[width=0.48\columnwidth]{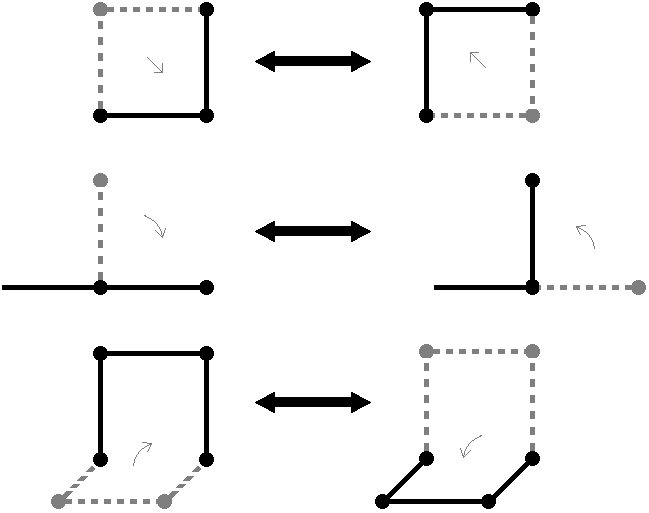}\\
    \caption{$n$-preserving Monte Carlo moves. For each of the three
      kind of moves, two reciprocal realizations are plotted.}
    \label{fig_mc_moves}
  \end{center}
\end{figure}

\subsection{Motion of the CM}
\label{sec_motion_cm}
In order to reproduce the diffusion dynamics of a polymer, the
algorithm is enriched by adding three types of $n$-preserving local
moves as shown in Figure~\ref{fig_mc_moves}.  These are: (i) the
one-vertex flip or kink move; (ii) the $90^\circ$ rotation of end
steps; (iii) the $90^\circ$ crankshaft move~\cite{Sokal_96}.  In all
cases, if the move leads to a violation of self-avoidance, the
attempted move is rejected. For a SAW of size $n$, each MCS consists
of one attempted Berretti-Sokal move followed by a sequence of $n$
attempts of moves of the kind (i)--(iii) with a random choice of the
location and of the specific $n$-preserving move. Since moves of kind
(iii) attempt to change the position of two vertices in place of that
of a single one, they are proposed with half the probability of moves
(i), (ii).  Note that for the whole set of three local moves to be
potentially applicable, a minimal number of $3$ steps is required; in
our simulations we then set $n_{\mathrm{min}}=4$ to have ensured
proper mobility at all sizes.
This minimal size is still sufficiently
small to guarantee the ergodicity of the
algorithm~\cite{berretti1985new}.
Ergodicity is in fact granted once it is assured the possibility of
passing from any $n$-step SAW to any other $n'$-step SAW
within a finite number of moves ($n_{\mathrm{min}}\le n,n'<\infty$).
In view of the reversibility
of the moves, this is equivalent to the capability of reaching
a specific target (say, e.g., a $4$-step SAW completely
straighten along the $x$ direction) from any $n$-step SAW.
It is easy to figure out a possible  finite sequence of moves performing the last
task as follows: first, the application of $n-4$ moves of the kind $-1$ reaches one of the
$726$~\cite{clisby2007enumeration,schram2011enumeration} distinct 
$4$-step SAWs occurring on the cubic lattice; then, within a finite sequence
of $n$-preserving moves the target is certainly reached. 
In all our simulations for the motion of CM we start
from a grand canonical equilibrium configuration.

Eventually, it is interesting to point out that the correct diffusive
properties for our polymer model are reproduced even neglecting
thermal fluctuations induced by the interaction with the solvent,
which might excite translational and rotational degrees of freedom of
the polymer. Such a simplification, enabling a larger statistical
sampling, is in line with kinetic Monte Carlo
standards~\cite{baumgartner1979,chern2001,burroughs2007}, and
is ultimately allowed by the fact that the overdamped motion of the CM is
in fact an attractor for a wide class of dynamical models.


\begin{figure}[t]
  \begin{center}
    \includegraphics[width=1.\columnwidth]{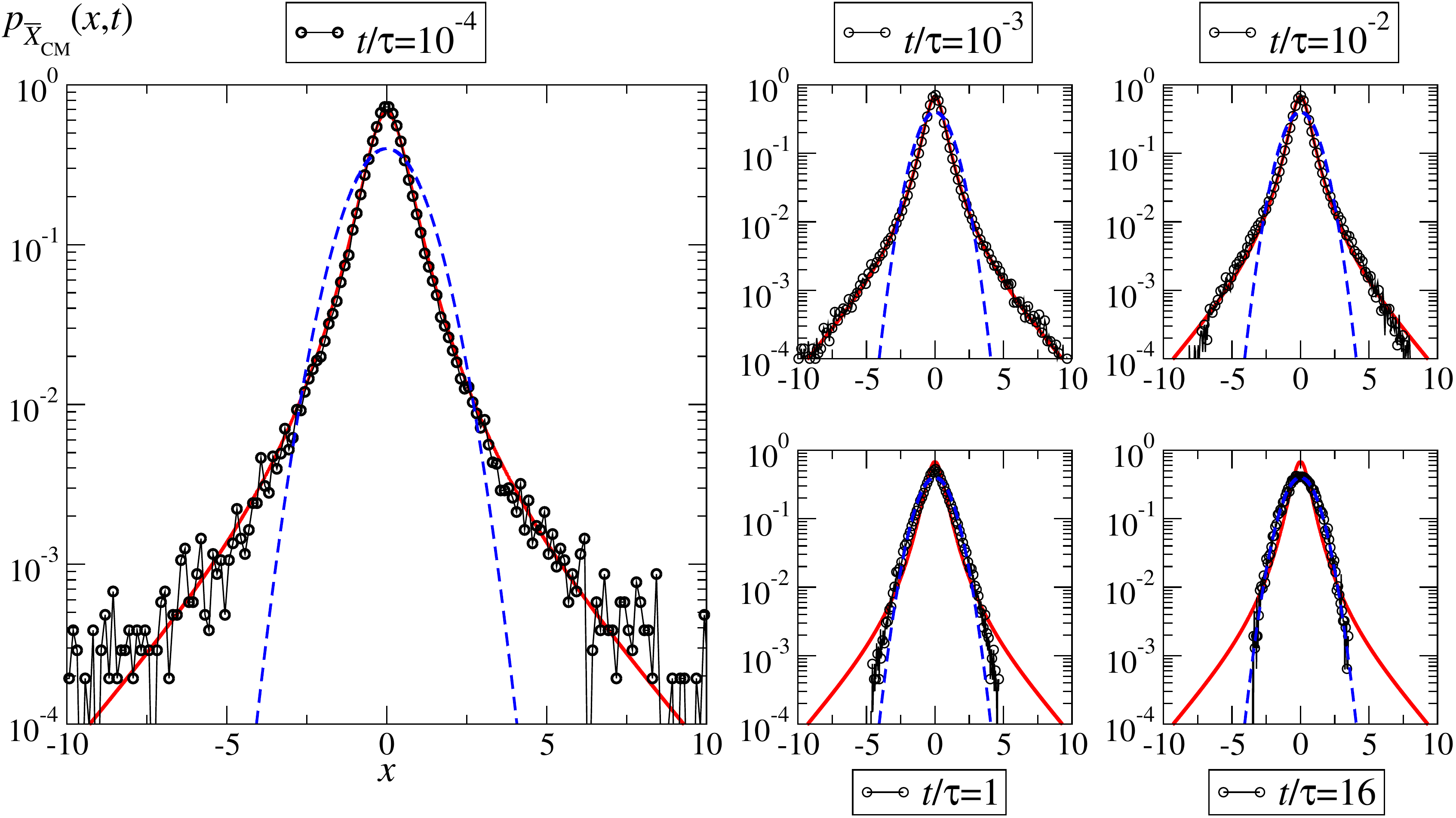}\\
    \caption{Non-Gaussian diffusion of SAWs. Exact theoretical
      predictions
      for the PDF of the unit-variance dimensionless variable
      $\overline{X}_{\mathrm{CM}}(t)\equiv
      X_{\mathrm{CM}}(t)/\sqrt{\mathbb{E}[X_{\mathrm{CM}}^2(t)]}$
      (curves) are compared
      with normalized displacement histograms obtained from Monte
      Carlo simulations (circles). With $t/\tau\leq10^{-2}$, histograms
      are built with $4.9\times10^5$ realizations; otherwise with
      $6.6\times10^4$ realizations. Full red lines are given by
      Eq.~\eqref{eq_initial_pdf}; dashed blue lines reproduce a zero-mean
      unit-variance Gaussian. } 
    \label{fig_p_x}
  \end{center}
\end{figure}

\section{Discussion}
According to
Eq.~\eqref{eq_initial_pdf} the shape of
the initial PDF $p_{\overline{X}_{\mathrm{CM}}}(x,0^+)$ is universal close to
criticality, being identified by the value of the fugacity,
by $\gamma$, and by the exponent of the $D$-vs-$n$ relation ($-1$ in
the present case).  The only remaining detail is $n_{\mathrm{min}}$,
fixed equal to $4$ in our simulations. Hence, no fitting parameter are
left free. Figure.~\ref{fig_p_x} highlights a striking agreement between
the universal predictions and the  normalized histograms for
$p_{\overline{X}_{\mathrm{CM}}}(x,t)$ obtained from the Monte Carlo simulations.
Initially, $p_{\overline{X}_{\mathrm{CM}}}$ sticks close to the PDF of the anomalous fixed
point $z=z_{\mathrm{c}}$ (characterized in the present universality class by an
infinite kurtosis~\cite{nampoothiri2022}), and only with
$t\simeq10^{-2}\tau$ some deviation develops at
probability values of the order of $10^{-4}$. As $t\simeq\tau$,
$p_{\overline{X}_{\mathrm{CM}}}$ crosses over  the (trivial) Gaussian
fixed-point. We remind the reader that, as $z\to z_{\mathrm{c}}^-$,
$\tau$ becomes infinite.
Although we concentrate here on the short- and long-time limit in
which analytical predictions are handy, using the Gillespie
algorithm~\cite{gillespie1977} for the process $N(t)$ in combination
with an ordinary Wiener process it is also possible to simulate
Eq.~\eqref{eq_langevin} and obtain thus the theoretical full time
evolution  of $p_{\overline{X}_{\mathrm{CM}}}(x,t)$. 

\begin{figure}[t]
  \begin{center}
    \includegraphics[width=0.5\columnwidth]{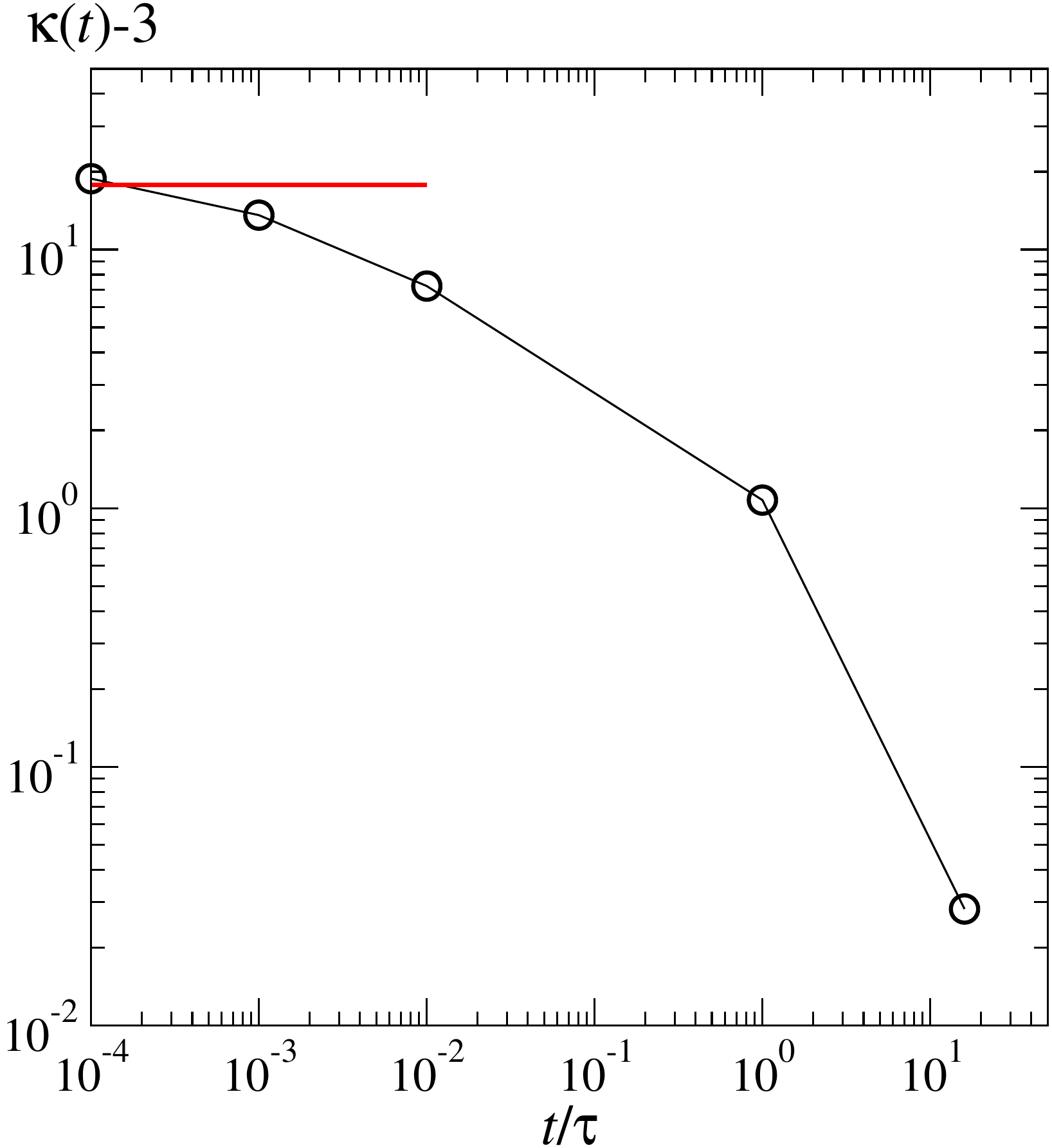}\\
    \caption{Time evolution for the excess kurtosis in Monte Carlo
      simulations (circles). The horizontal line represents the theoretical
      asymptotic behaviour, according to Eq.~\eqref{eq_initial_pdf}.} 
    \label{fig_kurtosis}
  \end{center}
\end{figure}

The crossover behaviour is made evident in
Figure~\ref{fig_kurtosis}, where it is reported the time evolution of
the excess kurtosis,
\begin{equation}
  \kappa(t)-3\equiv
  \dfrac{
    \mathbb{E}\left[\left(\overline{X}_{\mathrm{CM}}
      -\mathbb{E}\left[\overline{X}_{\mathrm{CM}}\right]\right)^4\right]
  }{
    \left(\mathbb{E}\left[\left(\overline{X}_{\mathrm{CM}}
      -\mathbb{E}\left[\overline{X}_{\mathrm{CM}}\right]\right)^2\right]\right)^2
  }
  -3
  =\dfrac{
    \mathbb{E}\left[\overline{X}_{\mathrm{CM}}^4\right]
  }{
    \left(\mathbb{E}\left[\overline{X}_{\mathrm{CM}}^2\right] \right)^2}
  -3
\end{equation}
according to the Monte Carlo simulations.
Despite the large sensitivity of the fourth moment of the PDF to
fluctuations in the tails, Fig.~\ref{fig_kurtosis} confirms a
substantial agreement between simulations and theoretical predictions
as $t\to0^-$, and highlights that the excess kurtosis tends to zero
(Gaussian PDF) above the autocorrelation time scale. 

Stochastic models are typically constructed 
assuming specific features for noise terms in
the equations. Here we followed a different path, in which the
model begets from a genuine microscopic instance endowed with an
autonomous dynamics.  
We have thus been able to provide the first evidence in which the
Brownian non-Gaussian diffusion compliant with a stochastic model is
independently and exactly produced by a microscopic dynamics.
Noting that the anomalous diffusion  of the polymer CM
is ruled by its equilibrium  critical
behaviour,  and ultimately by the value of the entropic critical
exponent $\gamma$, it  would be worthwhile to extend future theoretical
and numerical  considerations to models of polymers either
characterized by  different, fixed, topologies~\cite{duplantier1989}
(i.e. linear vs star or branched polymer )  or to  polymeric systems
still in diluted condition  but undergoing conformational phase
transitions such as the 
Theta- or the
adsorption-transition~\cite{deGennes1972,vanderzande1998}.
For instance,  
one should compare the theoretical predictions  of the CM diffusion of
linear polymers undergoing a collapse transition from a swollen to a
compact phase with gran canonical (i.e $N$-varying ensemble) Monte Carlo
simulations of   interacting self-avoiding walks  on the cubic lattice~\cite{vanderzande1998}.
Similarly, one might look at the CM diffusion of three-dimensional SAWs once adsorbed on 
an attractive impenetrable wall~\cite{van2015statistical}  or localised in proximity of 
an interface separating two bulk regions characterized by different solvent 
conditions~\cite{leibler1982theory,causo2003monte}.

\section*{Acknowledgments}
This work has
been partially supported by the University of Padova BIRD191017
project ``Topological statistical dynamics''.

\section*{References}
\bibliography{draft_bibliography}

\begin{thebibliography}{10}

\bibitem{wang2009}
Bo~Wang, Stephen~M Anthony, Sung~Chul Bae, and Steve Granick.
\newblock Anomalous yet brownian.
\newblock {\em Proceedings of the National Academy of Sciences},
  106(36):15160--15164, 2009.

\bibitem{wang2012}
Bo~Wang, James Kuo, Sung~Chul Bae, and Steve Granick.
\newblock When brownian diffusion is not gaussian.
\newblock {\em Nature materials}, 11(6):481, 2012.

\bibitem{toyota2011}
Toshihiro Toyota, David~A Head, Christoph~F Schmidt, and Daisuke Mizuno.
\newblock Non-gaussian athermal fluctuations in active gels.
\newblock {\em Soft Matter}, 7(7):3234--3239, 2011.

\bibitem{yu2013}
Changqian Yu, Juan Guan, Kejia Chen, Sung~Chul Bae, and Steve Granick.
\newblock Single-molecule observation of long jumps in polymer adsorption.
\newblock {\em ACS Nano}, 7(11):9735, 2013.

\bibitem{yu2014}
Changqian Yu and Steve Granick.
\newblock Revisiting polymer surface diﬀusion in the extreme case of strong
  adsorption.
\newblock {\em Langmuir}, 30:14538, 2014.

\bibitem{chakraborty2020}
Indrani Chakraborty and Yael Roichman.
\newblock Disorder-induced fickian, yet non-gaussian diffusion in heterogeneous
  media.
\newblock {\em Physical Review Research}, 2(2):022020, 2020.

\bibitem{weeks2000}
Eric~R Weeks, John~C Crocker, Andrew~C Levitt, Andrew Schofield, and David~A
  Weitz.
\newblock Three-dimensional direct imaging of structural relaxation near the
  colloidal glass transition.
\newblock {\em Science}, 287(5453):627--631, 2000.

\bibitem{wagner2017}
Caroline~E Wagner, Bradley~S Turner, Michael Rubinstein, Gareth~H McKinley, and
  Katharina Ribbeck.
\newblock A rheological study of the association and dynamics of muc5ac gels.
\newblock {\em Biomacromolecules}, 18(11):3654--3664, 2017.

\bibitem{jeon2016}
Jae-Hyung Jeon, Matti Javanainen, Hector Martinez-Seara, Ralf Metzler, and Ilpo
  Vattulainen.
\newblock Protein crowding in lipid bilayers gives rise to non-gaussian
  anomalous lateral diffusion of phospholipids and proteins.
\newblock {\em Physical Review X}, 6(2):021006, 2016.

\bibitem{yamamoto2017}
Eiji Yamamoto, Takuma Akimoto, Antreas~C Kalli, Kenji Yasuoka, and Mark~SP
  Sansom.
\newblock Dynamic interactions between a membrane binding protein and lipids
  induce fluctuating diffusivity.
\newblock {\em Science advances}, 3(1):e1601871, 2017.

\bibitem{stylianidou2014}
Stella Stylianidou, Nathan~J Kuwada, and Paul~A Wiggins.
\newblock Cytoplasmic dynamics reveals two modes of nucleoid-dependent
  mobility.
\newblock {\em Biophysical journal}, 107(11):2684--2692, 2014.

\bibitem{parry2014}
Bradley~R Parry, Ivan~V Surovtsev, Matthew~T Cabeen, Corey~S O’Hern, Eric~R
  Dufresne, and Christine Jacobs-Wagner.
\newblock The bacterial cytoplasm has glass-like properties and is fluidized by
  metabolic activity.
\newblock {\em Cell}, 156(1-2):183--194, 2014.

\bibitem{munder2016}
Matthias~Christoph Munder, Daniel Midtvedt, Titus Franzmann, Elisabeth Nuske,
  Oliver Otto, Maik Herbig, Elke Ulbricht, Paul M{\"u}ller, Anna Taubenberger,
  Shovamayee Maharana, et~al.
\newblock A ph-driven transition of the cytoplasm from a fluid-to a solid-like
  state promotes entry into dormancy.
\newblock {\em elife}, 5:e09347, 2016.

\bibitem{cherstvy2018}
Andrey~G Cherstvy, Oliver Nagel, Carsten Beta, and Ralf Metzler.
\newblock Non-gaussianity, population heterogeneity, and transient
  superdiffusion in the spreading dynamics of amoeboid cells.
\newblock {\em Physical Chemistry Chemical Physics}, 20(35):23034--23054, 2018.

\bibitem{li2019}
Yunyun Li, Fabio Marchesoni, Debajyoti Debnath, and Pulak~K Ghosh.
\newblock Non-gaussian normal diffusion in a fluctuating corrugated channel.
\newblock {\em Physical Review Research}, 1(3):033003, 2019.

\bibitem{cuetos2018}
Alejandro Cuetos, Neftal{\'\i} Morillo, and Alessandro Patti.
\newblock Fickian yet non-gaussian diffusion is not ubiquitous in soft matter.
\newblock {\em Physical Review E}, 98(4):042129, 2018.

\bibitem{hapca2008}
Simona Hapca, John~W Crawford, and Iain~M Young.
\newblock Anomalous diffusion of heterogeneous populations characterized by
  normal diffusion at the individual level.
\newblock {\em Journal of the Royal Society Interface}, 6(30):111--122, 2008.

\bibitem{pastore2021rapid}
Raffaele Pastore, Antonio Ciarlo, Giuseppe Pesce, Francesco Greco, and Antonio
  Sasso.
\newblock Rapid fickian yet non-gaussian diffusion after subdiffusion.
\newblock {\em Physical Review Letters}, 126(15):158003, 2021.

\bibitem{pastore2015}
Raﬀaele Pastore and Guido Raos.
\newblock Glassy dynamics of a polymer monolayer on a heterogeneous disordered
  substrate.
\newblock {\em Soft Matter}, 11:8083, 2015.

\bibitem{miotto2021length}
Jos{\'e}~M Miotto, Simone Pigolotti, Aleksei~V Chechkin, and S{\'a}ndalo
  Rold{\'a}n-Vargas.
\newblock Length scales in brownian yet non-gaussian dynamics.
\newblock {\em Physical Review X}, 11(3):031002, 2021.

\bibitem{pastore2022}
Francesco Rusciano, Raffaele Pastore, and Francesco Greco.
\newblock Fickian non-gaussian diffusion in glass-forming liquids.
\newblock {\em Physical Review Letters}, 128:168001, 2022.

\bibitem{beck2003}
Christian Beck and Ezechiel~GD Cohen.
\newblock Superstatistics.
\newblock {\em Physica A: Statistical mechanics and its applications},
  322:267--275, 2003.

\bibitem{beck2006}
Christian Beck.
\newblock Superstatistical brownian motion.
\newblock {\em Progress of Theoretical Physics Supplement}, 162:29--36, 2006.

\bibitem{chubynsky2014}
Mykyta~V Chubynsky and Gary~W Slater.
\newblock Diffusing diffusivity: a model for anomalous, yet brownian,
  diffusion.
\newblock {\em Physical review letters}, 113(9):098302, 2014.

\bibitem{chechkin2017}
Aleksei~V Chechkin, Flavio Seno, Ralf Metzler, and Igor~M Sokolov.
\newblock Brownian yet non-gaussian diffusion: from superstatistics to
  subordination of diffusing diffusivities.
\newblock {\em Physical Review X}, 7(2):021002, 2017.

\bibitem{jain2017}
Rohit Jain and KL~Sebastian.
\newblock Diffusing diffusivity: a new derivation and comparison with
  simulations.
\newblock {\em Journal of Chemical Sciences}, 129(7):929--937, 2017.

\bibitem{tyagi2017}
Neha Tyagi and Binny~J Cherayil.
\newblock Non-gaussian brownian diffusion in dynamically disordered thermal
  environments.
\newblock {\em The Journal of Physical Chemistry B}, 121(29):7204--7209, 2017.

\bibitem{miyaguchi2017}
Tomoshige Miyaguchi.
\newblock Elucidating fluctuating diffusivity in center-of-mass motion of
  polymer models with time-averaged mean-square-displacement tensor.
\newblock {\em Physical Review E}, 96(4):042501, 2017.

\bibitem{sposini2018}
Vittoria Sposini, Aleksei~V Chechkin, Flavio Seno, Gianni Pagnini, and Ralf
  Metzler.
\newblock Random diffusivity from stochastic equations: comparison of two
  models for brownian yet non-gaussian diffusion.
\newblock {\em New Journal of Physics}, 20(4):043044, 2018.

\bibitem{sposini2018first}
V~Sposini, A~Chechkin, and R~Metzler.
\newblock First passage statistics for diffusing diffusivity.
\newblock {\em Journal of Physics A: Mathematical and Theoretical},
  52(4):04LT01, 2018.

\bibitem{barkai2020}
E~Barkai and S~Burov.
\newblock Packets of diffusing particles exhibit universal exponential tails.
\newblock {\em Physical Review Letters}, 124:060603, 2020.

\bibitem{wang2020}
W~Wang, E~Barkai, and S~Burov.
\newblock Packets of diffusing particles exhibit universal exponential tails.
\newblock {\em Entropy}, 22:697, 2020.

\bibitem{sokolov2021ld}
A~Pacheco-Pozo and I~M Sokolov.
\newblock Large deviations in continuous-time random walks.
\newblock {\em Physical Review E}, 103:042116, 2021.

\bibitem{sokolov2021}
A~Pacheco-Pozo and I~M Sokolov.
\newblock Convergence to a gaussian by narrowing of central peak in brownianyet
  non-gaussian diffusion in disordered environments.
\newblock {\em Physical Review Letters}, 127:120601, 2021.

\bibitem{baldovin2019}
F~Baldovin, E~Orlandini, and F~Seno.
\newblock Polymerization induces non-gaussian diffusion.
\newblock {\em Frontiers in Physics}, 7:124, 2019.

\bibitem{nampoothiri2021}
S~Nampoothiri, E~Orlandini, F~Seno, and F~Baldovin.
\newblock Polymers critical point originates brownian non-gaussian diffusion.
\newblock {\em Physical Review E}, 104:L062501, 2021.

\bibitem{nampoothiri2022}
S~Nampoothiri, E~Orlandini, F~Seno, and F~Baldovin.
\newblock Brownian non-gaussian polymer diffusion and queuing theory in the
  mean-field limit.
\newblock {\em New J. Phys.}, 24:023003, 2022.

\bibitem{hidalgo2020}
M~Hidalgo-Soria and E~Barkai.
\newblock Hitchhiker model for laplace diffusion processes in the cell
  environment.
\newblock {\em Physical Review E}, 102:012109, 2020.

\bibitem{deGennes1972}
P-G de~Gennes.
\newblock Exponents for the excluded volume problem as derived by the wilson
  method.
\newblock {\em Physics Letters A}, 38:339–340, 1972.

\bibitem{deGennes1979}
P-G de~Gennes.
\newblock {\em Scaling Concepts in Polymer Physics}.
\newblock Cornell University Press, 1979.

\bibitem{vanderzande1998}
C~Vanderzande.
\newblock {\em Lattice Models of Polymers}.
\newblock Cambridge University Press, 1998.

\bibitem{madras2013}
N~Madras and G~Slade.
\newblock {\em The Self-Avoiding Walk}.
\newblock Springer, 2013.

\bibitem{baumgartner1979}
A.~Baumg\"artner and K.~Binder.
\newblock Monte carlo studies on the freely jointed polymer chain with excluded
  volume interaction.
\newblock {\em J. Chem. Phys.}, 71:2541, 1979.

\bibitem{chern2001}
Shyh-Shi Chern, Alfredo~E. C\'ardenas, and Rob~D. Coalson.
\newblock Three-dimensional dynamic monte carlo simulations of driven polymer
  transport through a hole in a wall.
\newblock {\em J. Chem. Phys.}, 115:7772, 2001.

\bibitem{burroughs2007}
N.~J. Burroughs and D.~Marenduzzo.
\newblock Nonequilibrium-driven motion in actin networks: Comet tails and
  moving beads.
\newblock {\em Phys. Rev. Lett.}, 98:238302, 2007.

\bibitem{clisby2013calculation}
N~Clisby.
\newblock Calculation of the connective constant for self-avoiding walks via
  the pivot algorithm.
\newblock {\em Journal of Physics A: Mathematical and Theoretical},
  46(24):245001, 2013.

\bibitem{clisby2017scale}
N~Clisby.
\newblock Scale-free monte carlo method for calculating the critical exponent
  $\gamma$ of self-avoiding walks.
\newblock {\em Journal of Physics A: Mathematical and Theoretical}, 50:264003,
  2017.

\bibitem{Doi1992}
M~Doi and Edwards~S F.
\newblock {\em The Theory of Polymer Dynamics}.
\newblock Oxford University Press, 1992.

\bibitem{Feller1968}
W~Feller.
\newblock {\em An Introduction to Probability Theory and Its Applications}.
\newblock John Wiley \& Sons, 1968.

\bibitem{bochner2020harmonic}
Salomon Bochner.
\newblock {\em Harmonic analysis and the theory of probability}.
\newblock University of California press, 2020.

\bibitem{berretti1985new}
A~Berretti and A~D Sokal.
\newblock New monte carlo method for the self-avoiding walk.
\newblock {\em Journal of Statistical Physics}, 40:483, 1985.

\bibitem{Sokal_96}
A~D Sokal.
\newblock Monte carlo methods for the self avoiding walk.
\newblock {\em Nuc.Phys. B - Proceeding Supplements}, 47:172--179, 1996.

\bibitem{clisby2007enumeration}
N~Clisby, R~Liang, and G~Slade.
\newblock Self-avoiding walk enumeration via the lace expansion.
\newblock {\em Journal of Physics A: Mathematical and Theoretical}, 40:10973,
  2007.

\bibitem{schram2011enumeration}
D~Schram, G~T Barkema, and R~H Bisseling.
\newblock Exact enumeration of self-avoiding walks.
\newblock {\em J. Stat. Mech.}, page P06019, 2011.

\bibitem{gillespie1977}
Daniel~T Gillespie.
\newblock Exact stochastic simulation of coupled chemical reactions.
\newblock {\em Journal of Physical Chemistry}, 81:2340–2361, 1977.

\bibitem{duplantier1989}
B~Duplantier.
\newblock Statistical mechanics of polymer networks of any topology.
\newblock {\em Journal of Statistical Physics}, 54(3/4):581, 1989.

\bibitem{van2015statistical}
EJ~Janse Van~Rensburg.
\newblock {\em The statistical mechanics of interacting walks, polygons,
  animals and vesicles}.
\newblock Oxford Lecture Mathematics and, 2015.

\bibitem{leibler1982theory}
Ludwik Leibler.
\newblock Theory of phase equilibria in mixtures of copolymers and
  homopolymers. 2. interfaces near the consolute point.
\newblock {\em Macromolecules}, 15(5):1283--1290, 1982.

\bibitem{causo2003monte}
Maria~Serena Causo and Stuart~G Whittington.
\newblock A monte carlo investigation of the localization transition in random
  copolymers at an interface.
\newblock {\em Journal of Physics A: Mathematical and General}, 36(13):L189,
  2003.

\end{thebibliography}

\end{document}